\documentclass[twocolumn]{aastex63}
\accepted{for publication in ApJ \today}
\shorttitle{}
\shortauthors{Lee et al.}
\begin{document}
\title{
Further evidence for significant luminosity evolution in supernova cosmology
}

\correspondingauthor{Young-Wook Lee, Chul Chung}
\email{ywlee2@yonsei.ac.kr, chulchung@yonsei.ac.kr}

\author[0000-0002-2210-1238]{Young-Wook Lee}
\affil{Department of Astronomy, Yonsei University, Seoul 03722, Republic of Korea}
\affil{Center for Galaxy Evolution Research, Yonsei University, Seoul 03722, Republic of Korea}
%\affil{These authors contributed equally to this paper.}

\author[0000-0001-6812-4542]{Chul Chung}
\affil{Department of Astronomy, Yonsei University, Seoul 03722, Republic of Korea}
\affil{Center for Galaxy Evolution Research, Yonsei University, Seoul 03722, Republic of Korea}
%\affil{These authors contributed equally to this paper.}
%
%
\author[0000-0002-5261-5803]{Yijung Kang}
\affil{
Gemini Observatory/NSF’s NOIRLab, Casilla 603, La Serena, Chile
}
\author[0000-0002-5751-3697]{M. James Jee}
\affil{Department of Astronomy, Yonsei University, Seoul 03722, Republic of Korea}
\affil{Department of Physics, University of California, Davis, One Shields Avenue, Davis, CA 95616, USA}

\begin{abstract}
Supernova (SN) cosmology is based on the assumption that the corrected luminosity of SN Ia would not evolve with redshift. 
Recently, our age dating of stellar populations in early-type host galaxies (ETGs) from high-quality spectra has shown that this key assumption is most likely in error. 
It has been argued though that the age-Hubble residual (HR) correlation from ETGs is not confirmed from two independent age datasets measured from multi-band optical photometry of host galaxies of all morphological types. 
Here we show, however, that one of them is based on highly uncertain and inappropriate luminosity-weighted ages derived, in many cases, under serious template mismatch. 
The other dataset employs more reliable mass-weighted ages, but the statistical analysis involved is affected by regression dilution bias, severely underestimating both the slope and significance of the age-HR correlation. 
Remarkably, when we apply regression analysis with a standard posterior sampling method to this dataset comprising a large sample ($N=102$) of host galaxies, very significant ($> 99.99 \%$) correlation is obtained between the global population age and HR with the slope ($-0.047 \pm 0.011$~mag/Gyr) highly consistent with our previous spectroscopic result from ETGs. 
For the local age of the environment around the site of SN, a similarly significant ($> 99.96 \%$) correlation is obtained with a steeper slope ($-0.057 \pm 0.016$~mag/Gyr). 
Therefore, the SN luminosity evolution is strongly supported by the age dating based on multi-band optical photometry and can be a serious systematic bias in SN cosmology.
\end{abstract}


\section{Introduction}
\label{s1}

The inference of dark energy in supernova (SN) cosmology is based on the assumption that the SN luminosity, after the empirical standardization, would not evolve with redshift \citep{1998AJ....116.1009R, 1998ApJ...507...46S, 1999ApJ...517..565P}. 
As recognized by early investigators (see Figure~3 of \citealt{1998ApJ...507...46S}; see also \citealt{1998AJ....116.1009R} and \citealt{1999ApJ...517..565P}), this key assumption can be best tested at low-$z$ by looking for any correlation between the population age of a host galaxy and the Hubble residual (HR) of SN. 
While the correlations between HR and host galaxy properties, such as stellar mass and star formation rate, are now well established \citep[e.g.,][]{2010ApJ...715..743K, 2010MNRAS.406..782S, 2015ApJ...802...20R, 2018ApJ...854...24K}, there is, however, a paucity of literature on robust measurements of stellar population ages for host galaxies. 
Recently, \citet{2020ApJ...889....8K} have obtained the direct and reliable estimates of population ages for a sample of local early-type host galaxies (ETGs) from exceptionally high-quality (signal-to-noise ratio $\sim 175$) spectra. 
Based on this new age dataset, we found a correlation between population age and HR which indicates a non-negligible luminosity evolution in SN cosmology. 
While this result is based on a sample of ETGs, there is no theoretical reason that the age-HR correlation observed in ETGs should not extend to other types of host galaxies. 
Nevertheless, since type Ia SNe are discovered in all morphological types of galaxies, it is important to check whether this correlation is confirmed by a larger sample of host galaxies comprising all morphological types.

\citet{2020ApJ...896L...4R} have claimed, however, that this age-HR slope obtained from ETGs is not confirmed from the two independent age datasets measured from multi-band optical photometry of host galaxies of all morphological types. 
Based on this result, they argued that there is no evidence for SN Ia luminosity evolution. This on-going debate further underscores that the age-HR slope would determine the significance of the luminosity evolution and, therefore, the validity of the key assumption in SN cosmology. 
Because of its important implication for the inference of dark energy from SN cosmology, the origin of this apparent disparity between \citet{2020ApJ...889....8K} and \citet{2020ApJ...896L...4R} must be investigated thoroughly. 
The purpose of this paper is to show that, when the regression analysis of the \citet{2020ApJ...896L...4R} dataset is performed in a consistent and standard manner, very significant age-HR correlation is also obtained from a large sample of host galaxies comprising all morphological types with the slope highly consistent with our previous spectroscopic result from ETGs.

\begin{figure*}[htbp]
\centering
\includegraphics[angle=-90,scale=0.8]{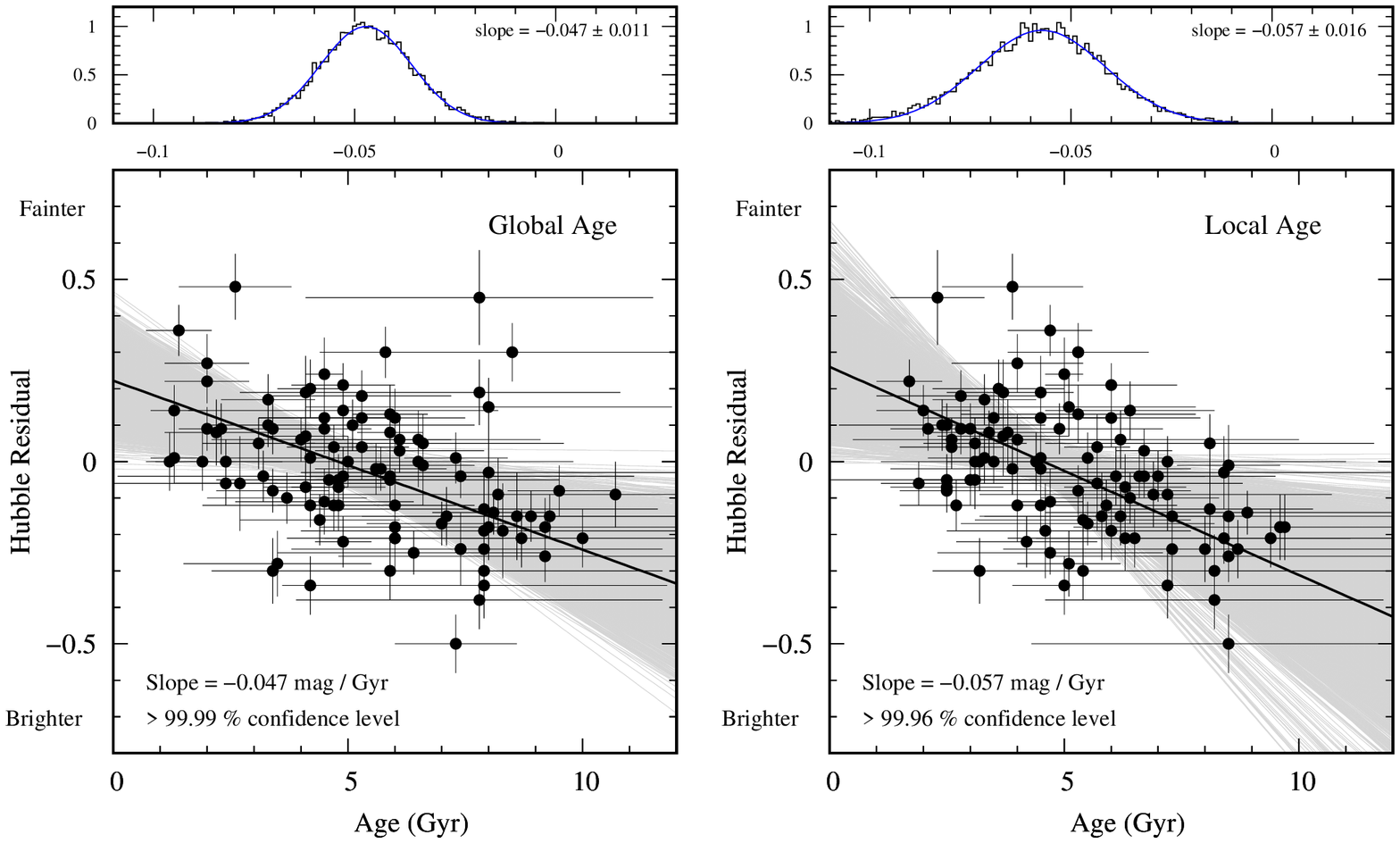}
\caption{
The strong correlation between population age and HR for host galaxies of all morphological types (data from \citealt{2019ApJ...874...32R}). 
The left panel is for the mass-weighted global ages of 102 host galaxies. 
The solid line is the best regression fit obtained from MCMC posterior sampling method, which shows a very significant ($4.3 \sigma$) correlation between age and HR with the slope in excellent agreement with our previous spectroscopic result from ETGs. 
The right panel is for the local age of the environment around the site of SN in a host galaxy. 
Again, similarly significant ($3.6 \sigma$) correlation is obtained with an even steeper slope ($-0.057$~mag/Gyr).
\label{f1}}
\end{figure*}

\section{Re-examining stellar population ages from multi-band optical photometry
}
\label{s2}

One of the two age datasets employed by \citet[][their Figure~3]{2020ApJ...896L...4R} is originated from low-$z$ host galaxy sample of \citet{2018ApJ...867..108J}. 
\citet{2018ApJ...867..108J} used the Pan-STARRS $grizy$ and Sloan Digital Sky Survey (SDSS) $u$~band  photometry together with the code Z-PEG \citep{2002A&A...386..446L}, which was originally designed to estimate photometric redshift, but can also be used to derive other parameters including luminosity-weighted age, if properly employed. 
However, neither \citet{2020ApJ...896L...4R} nor \citet{2018ApJ...867..108J} provide the age data and their uncertainties. 
Without the crucial error bars for ages in Figure~3 of \citet{2020ApJ...896L...4R}, it is impossible to assess the validity of their ages and the statistical significance of their claim derived from these ages. 

Therefore, in an effort to investigate the reliability of their ages, we have selected 13 ETGs by cross-matching the \citet{2018ApJ...867..108J} sample with the \citet{2020ApJ...889....8K} ETG sample for which reliable estimates for ages are available from high-quality spectra. 
For this ETG subsample, we have re-enacted the procedures adopted by \citet{2018ApJ...867..108J} by deriving ages using Z-PEG and the same $ugrizy$ photometric data. 
The redshift and morphological classification for this low-$z$ sample was adopted from the NASA Extragalactic Database as listed in \citet{2020ApJ...889....8K}. 
Out of these 13 ETGs in common with \citet{2018ApJ...867..108J}, we found catastrophic spectral energy distribution (SED) template mismatches ($\chi^2 = 20$ -- 254; RMS error $\approx$ 1.15~mag) for 6 galaxies, and therefore the derived ages of these galaxies should be highly uncertain, if not meaningless. The origin for this mismatch is not clearly identified, seriously questioning the validity of ages for a significant fraction of galaxies in the \citet{2018ApJ...867..108J} sample. 
For the remaining 7 ETGs, we obtained the ages with $\chi^2 < 20$ (RMS error $\approx$ 0.11~mag), but the Z-PEG derived ages are still underestimated by $\sim$3~Gyr compared to the spectroscopic ages derived by \citet{2020ApJ...889....8K}, illustrating the well-known limitation of the luminosity-weighted ages from multi-band optical photometry \citep[see, e.g.,][]{2007ApJ...664..215L, 2011Ap&SS.331....1W}. 

For galaxies with on-going or recent star formation (most cases in the \citealt{2018ApJ...867..108J} sample), the luminosity-weighted age derived from the photometric SED would be further biased toward the younger age. 
This is because even a small fraction of very young stars in a galaxy can significantly affect its SED \citep[see][]{2007ApJ...664..215L, 2011ApJ...740...92G}. 
The majority of stellar populations in such galaxies can still be markedly older than the determined mean age. 
That the ages of \citet{2018ApJ...867..108J} are highly uncertain and underestimated can also be assessed from a severe internal inconsistency in \citet{2020ApJ...896L...4R} between their Figures~2 and 3. 
Figure~2 of \citet{2020ApJ...896L...4R} shows the age distribution of host galaxies based on more reliable mass-weighted ages of \citet{2019ApJ...874...32R}, which has a mean of $\sim$5~Gyr at $z \sim 0.14$ ($\sim$5.7~Gyr at $z = 0.0$). 
This should be compared to the age distribution in their Figure~3 based on the \citet{2018ApJ...867..108J} dataset, which has a mean of only $\sim$2.3~Gyr at the local universe. 
Therefore, when the population age is derived from photometric SED, the luminosity-weighted age is not appropriate for the present study requiring the true average age of stellar populations. 
Instead, we need carefully measured mass-weighted age which is more relevant to the SN progenitor age in a host galaxy \citep[see][]{2011ApJ...740...92G, 2019ApJ...874...32R}.

In addition to these critical problems in their ages, the HRs in \citet{2018ApJ...867..108J} further include the host-mass correction. 
In the analysis for the age-HR correlation, this is a very inappropriate treatment because the host mass is most likely a proxy for the population age. 
\citet[][see their Figure~9]{2020ApJ...889....8K} found a very tight ($>99.99 \%$) correlation between host-mass and age from high-quality spectra for early-type host galaxies, while they found no correlation with metallicity at a similar mass range where \citet{2010ApJ...715..743K} and \citet{2013ApJ...770..108C} found the correlation between host mass and HR. 
A similar correlation between galaxy mass and population age was also reported by \citet{2018NatAs...2..483V} for a large sample of non-host galaxies. 
Because of this correlation, applying the host-mass correction by itself would further undermine the correlation between age and HR.\footnote{ 
At given redshift, this empirical correction for host mass can indeed reduce the scatter in HR. 
However, since the redshift evolution of host mass is small for the redshift range ($z < 1 - 1.3$) relevant to SN cosmology, this empirical treatment, unlike the direct correction based on age (Figure~16 of \citealt{2020ApJ...889....8K}), has no impact on cosmology \citep[consistent with a zero slope; see Figure~13 of][]{2014A&A...568A..22B}. 
Therefore, the current practice of using a correction based on host mass cannot correct for the SN luminosity evolution with redshift.}
It is therefore not surprising to see that the correlation between age and HR is smeared out in Figure~3 of \citet{2020ApJ...896L...4R} by using this problematic dataset.

\section{Correlation between age and Hubble residual from all types of host galaxies}
\label{s3}

In order to overcome the problems in age dating from photometric SED, \citet{2019ApJ...874...32R} have devised a clever and efficient technique for measuring the mass-weighted age, which can provide more reliable average age of stellar populations in a host galaxy. 
Their technique is based on a Markov chain Monte Carlo (MCMC) sampling method to determine the most probable star formation history (SFH), which was then implemented in the updated version of the population synthesis model of \citet{2010ApJ...712..833C}. 
As such, the \citet{2019ApJ...874...32R} age dating is a significant improvement over a similar age dataset of \citet{2011ApJ...740...92G}.
Using their technique applied to SDSS $ugriz$ photometric SED, \citet{2019ApJ...874...32R} have measured, with adequate accuracy, mass-weighted ages for 102 host galaxies of all morphological types in $0.05 < z < 0.2$. 

Figure~\ref{f1} shows this dataset for population age and HR from \citet{2019ApJ...874...32R} both for the global age of a host galaxy and for the local age measured in the vicinity (1.5 -- 3~kpc radius) of the SN Ia site.
\citet{2019ApJ...874...32R} used the SN sample of \citet{2013ApJ...763...88C} for the HR information.
Since the \citet{2019ApJ...874...32R} sample is confined to a narrow redshift range, the effect of redshift evolution is negligible within their sample.
To properly account for both measurement errors and intrinsic scatter in the regression analysis, the MCMC posterior sampling method implemented in the LINMIX package \citep{2007ApJ...665.1489K} is most commonly used in SN host galaxy studies \citep[e.g.,][]{2010ApJ...715..743K, 2011ApJ...740...92G, 2014MNRAS.438.1391P, 2020MNRAS.491.5897P} including our previous investigation for ETGs \citep{2020ApJ...889....8K}. 
\citet{2007ApJ...665.1489K} has shown that this maximum-likelihood estimator based on the Gaussian mixture model outperforms other common estimators and provides the least biased result for the regression analysis.\footnote{Nevertheless, the method implicitly assumes that the parent distribution of the independent variable follows a Gaussian mixture model.} 
Surprisingly, unlike the argument of \citet{2020ApJ...896L...4R}, when we apply this standard regression analysis method to the \citet{2019ApJ...874...32R} dataset comprising a large sample of host galaxies, very significant ($> 99.99 \%$) correlation is obtained between the global population age and HR with the slope ($-0.047 \pm 0.011$~mag/Gyr) in excellent agreement with the result ($-0.051 \pm 0.022$~mag/Gyr) of \citet{2020ApJ...889....8K} from high-quality spectroscopy of ETGs. 
\citet{2019ApJ...874...32R} suggested that this correlation might be more consistent with a step of $\sim$0.1~mag in the HR at an age of $\sim$8~Gyr, but we obtain more or less the same slope ($-0.054\pm 0.015$~mag/Gyr) even if we restrict the sample to host galaxies younger than 8~Gyr.
This indicates that, unlike the star formation rate-HR correlation \citep{2013A&A...560A..66R}, the potential effect of a nonlinearity is not significant in the age-HR correlation.
While the global age of a host galaxy can be used to infer the SN progenitor age, the local age around the SN Ia site would serve as a better proxy for the SN progenitor age. 
The right panel of Figure~\ref{f1} shows that a similarly significant ($> 99.96 \%$) correlation is also obtained between the local population age and HR with the slope ($-0.057 \pm 0.016$~mag/Gyr) again consistent with the slope of \citet{2020ApJ...889....8K}. 
Therefore, our previous result based on a small sample ($N = 34$) of ETGs is now confirmed from a large sample ($N = 102$) of host galaxies comprising all morphological types.

\begin{figure}[htbp]
\centering
\includegraphics[keepaspectratio,width=\textwidth,height=0.75\textheight]{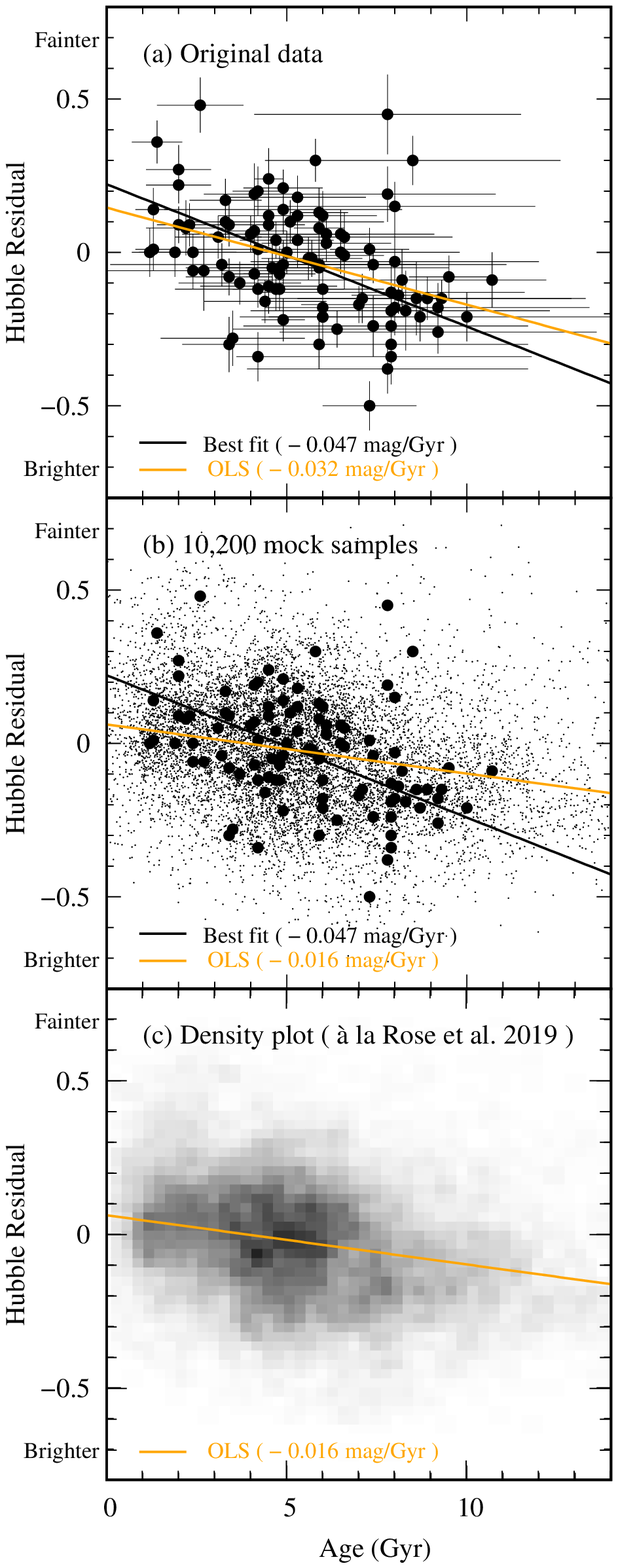}
\caption{
Regression dilution bias in \citet{2019ApJ...874...32R} analysis. 
(a) The ordinary least-squares (OLS) fit to 102 original data points already underestimates the slope compared to our best fit because of the measurement errors in the independent variable. 
(b) Generating 100 random mock samples (small dots) around each data point  leads to further attenuation of the slope. 
(c) A probability density plot based on these 10,200 mock data, as presented in \citet{2019ApJ...874...32R}, is significantly diluted, severely underestimating both the slope and significance of the age-HR correlation.
\label{f2}}
\end{figure}

The main argument of \citet{2020ApJ...896L...4R} is also based on this same dataset from \citet{2019ApJ...874...32R}, but they reached a very different conclusion for the slope much shallower than the one reported by \citet{2020ApJ...889....8K}. 
In order to understand the origin of this apparent disparity, we have carefully followed the procedures adopted by \citet{2019ApJ...874...32R} for which Figure~2 of \citet{2020ApJ...896L...4R} is based on. 
Figure~\ref{f2} shows our reproduction of their procedures. 
Unusually, Figure~2 of \citet{2020ApJ...896L...4R} only presents a probability density plot without showing the original individual 102 data points with error bars. 
Their density plot is based on a Monte-Carlo resampling method by generating 100 random mock samples around each data point according to the measurement errors. 
In doing so, however, the age range has been substantially stretched and, therefore, the slope obtained from the ordinary least-squares (OLS) fitting has been severely underestimated. 
This is the well-known regression dilution bias, which arises as a consequence of the measurement error in the independent variable and leads to the attenuation of both the regression slope and significance of the correlation \citep[see, e.g.,][]{2007ApJ...665.1489K}. 
Particularly, in the case of \citet{2019ApJ...874...32R, 2020ApJ...896L...4R} analysis, this effect has been doubled because the generation of the mock data stretches the distribution more horizontally than vertically\footnote{The mean error ($\sim$1.9~Gyr) for the age is $\sim$20\% of the interval ($\sim$9.5~Gyr) whereas the mean error ($\sim$0.079~mag) for the HR is $\sim$8\% of its interval ($\sim$0.98~mag).} and, more importantly, the OLS does not take into account the measurement errors of the mock data in the independent variable. 
In Figure~\ref{f2} we reproduce this double dilution bias that happened in \citet{2019ApJ...874...32R, 2020ApJ...896L...4R}. 
Since the public dataset of \citet{2019ApJ...874...32R} does not provide non-Gaussian error bars, we assume Gaussian errors here for mock data generation. 
However, as our experiment shows, the difference due to this non-Gaussianity is insignificant.

\begin{figure}[htbp]
\centering
\includegraphics[angle=-90,scale=0.8]{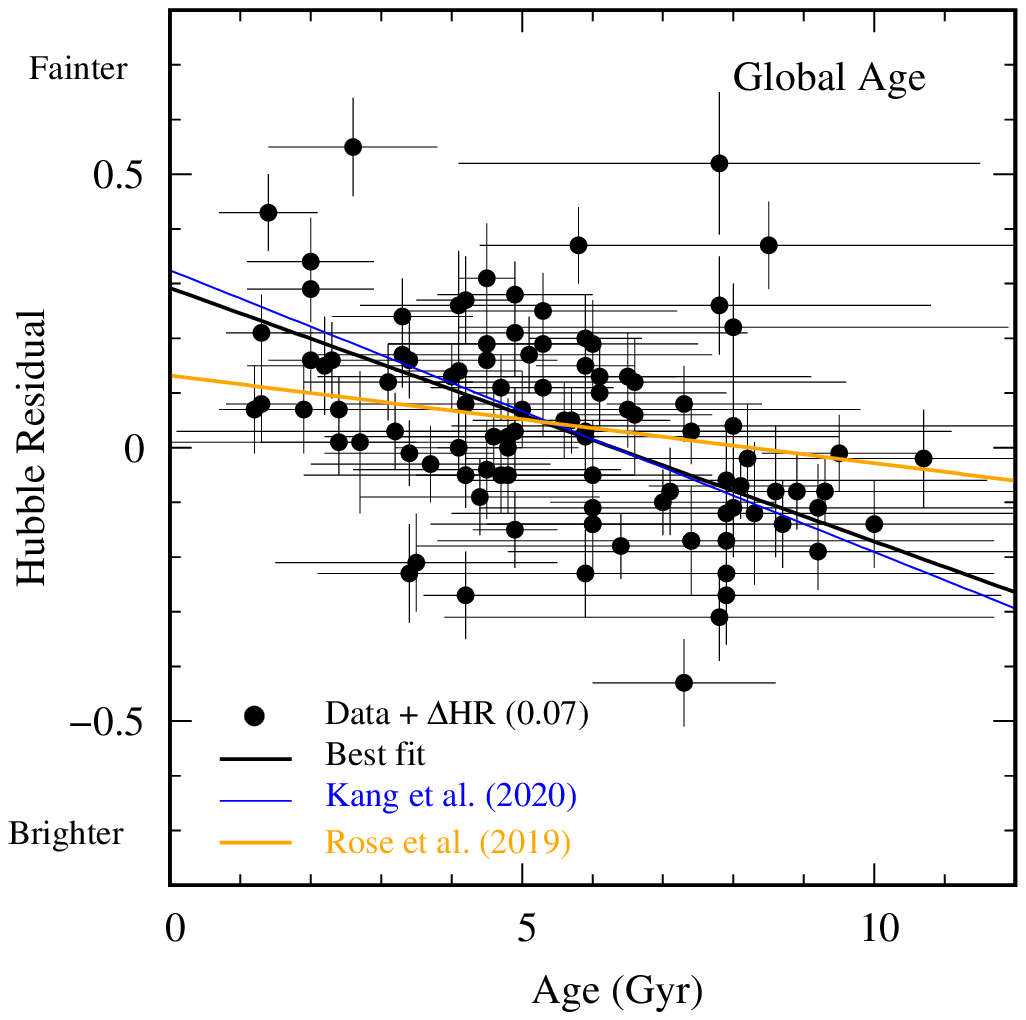}
\caption{
Comparison of the slopes. 
An HR offset of 0.07 mag is added to \citet{2019ApJ...874...32R} data to account for the difference in median redshift between the \citet[][$z \sim 0.04$]{2020ApJ...889....8K} and \citet[][$z \sim 0.14$]{2019ApJ...874...32R} samples. 
While the best fits obtained from the standard MCMC analysis for both \citet{2019ApJ...874...32R} and \citet{2020ApJ...889....8K} samples show excellent agreement each other, the analysis of \citet{2019ApJ...874...32R} is severely underestimating the slope with the regression line that does not represent the actual data points.
\label{f3}}
\end{figure}

Figure~\ref{f3} compares the slope obtained by us for the \citet{2019ApJ...874...32R} dataset comprising all types of host galaxies with that of \citet{2020ApJ...889....8K} for ETGs. 
Also compared is the slope reported by \citet{2019ApJ...874...32R}. 
A small HR shift of 0.07 mag is applied here to \citet{2019ApJ...874...32R} data to account for the difference in median redshift between the \citet[][$z \sim 0.04$]{2020ApJ...889....8K} and \citet[][$z \sim 0.14$]{2019ApJ...874...32R} samples. 
After this correction, the HR values would be equivalent to those calculated with respect to the cosmological model without $\Lambda$ ($\Omega_M = 0.27$, $\Omega_{\Lambda} = 0.00$). 
It is clear from this comparison that, while the slopes obtained from the standard MCMC posterior sampling method for both \citet{2020ApJ...889....8K} and \citet{2019ApJ...874...32R} samples show an excellent agreement with each other, the method of \citet{2019ApJ...874...32R} severely underestimates the slope because of the dilution bias.\footnote{When the regression line is obtained with the ``FITEXY'' estimator \citep{1992nrfa.book.....P}, even steeper slope ($-0.078$~mag/Gyr) is obtained, but \citet{2007ApJ...665.1489K} has shown that the FITEXY estimator is biased away from zero, while the OLS estimator is biased toward zero.} 
Note that most data points for ages older than $\sim$7~Gyr are placed below the regression line of \citet{2020ApJ...896L...4R}, illustrating that their regression does not fairly represent the distribution of actual data points. 
Their scientific conclusion (no luminosity evolution in SN cosmology) based on this problematic method is therefore seriously flawed.

\section{discussion}
\label{s4}

While the result of \citet{2020ApJ...889....8K} is based on the most direct population ages ever obtained for host galaxies from extremely high-quality spectra, it is limited to a small sample of ETGs. 
The present result is based on the ages derived from SED fitting of \citet{2019ApJ...874...32R} which are not as precise as those measured from spectral features, but a larger sample size coupled with adequate age accuracy have provided a far more significant ($>99.99 \%$, $4.3 \sigma$) correlation between population age and HR. 
Importantly, this result is no longer limited to ETGs but is based on host galaxies of all morphological types. 
Furthermore, unlike the \citet{2020ApJ...889....8K} analysis, no extrapolation in age is now required below 2.5~Gyr, because the \citet{2019ApJ...874...32R} sample contains younger host galaxies. 
In addition to the global age of a host galaxy, this study also presents the local population age around the site of SN, which is more relevant to the SN progenitor age\footnote{The SN progenitor age can be obtained by convolving a SFH of a host galaxy with the delay time distribution (DTD) of SN Ia \citep{2014MNRAS.445.1898C}. 
The difference between the population age and SN progenitor age can be estimated by employing SFHs and DTD in Figure~3 of \citet{2014MNRAS.445.1898C}. 
For the \citet{2019ApJ...874...32R} sample, we confirm that, on average, progenitor age is younger than population age by $\sim$1.3~Gyr, but this difference is larger at older ages and smaller at younger ages. 
Therefore, the slope in Figure~\ref{f1} would be somewhat ($\sim$15\%) steeper if we had used progenitor ages instead of population ages. 
While more detailed analysis requires a specific SFH for each host galaxy, this assures that the effect of the correction to SN progenitor age would be small in the derivation of the $\Delta$HR/$\Delta$age slope.}, and therefore is not strongly affected by the possible difference between the global and local population ages within a host galaxy. 
In these respects, the present result provides an independent confirmation for and a significant improvement over the result of \citet{2020ApJ...889....8K}. 
Therefore, the luminosity evolution stands up to scrutiny as a serious systematic bias in SN cosmology.

\acknowledgments

We thank Damien Le Borgne for his comments on the use of the Z-PEG code. 
We also thank the anonymous referees for a number of helpful comments and suggestions.
Support for this work was provided by the National Research Foundation of Korea (2017R1A2B3002919, 2017R1A5A1070354, 2017R1A2B2004644, \& 2020R1A4A2002885). 
The work of Y.K. is supported by the international Gemini Observatory, a program of NSF's NOIRLab, which is managed by the Association of Universities for Research in Astronomy (AURA) under a cooperative agreement with the National Science Foundation, on behalf of the Gemini partnership of Argentina, Brazil, Canada, Chile, the Republic of Korea, and the United States of America.

\bibliographystyle{aasjournal}

\end{document}